\documentclass[manuscript,screen]{acmart}
\AtBeginDocument{%
  }

\setcopyright{cc}
\copyrightyear{2026}
\acmYear{2026}
\acmDOI{}

\acmISBN{978-1-4503-XXXX-X/2018/06}

\begin{document}

\title[Understanding the Emerging Role of Trusted Flaggers under the EU Digital Services Act]{``There is literally zero funding'': Understanding the Emerging Role of Trusted Flaggers under the EU Digital Services Act}

\author{Marie-Therese Sekwenz}
\authornote{All authors contributed equally to this research.}
\affiliation{
    \institution{TU Delft}
    \country{Netherlands}
}
\author{Kyle Beadle}
\affiliation{
    \institution{UCL}
    \country{United Kingdom}
}
\author{Simon Parkin}
\affiliation{
    \institution{TU Delft}
    \country{Netherlands}
}


\renewcommand{\shortauthors}{Sekwenz et al.}

\begin{abstract}

The European Union's Digital Services Act (DSA) introduced regulatory mechanisms which serve as a way to manage harmful content online.
The recognition of Trusted Flaggers (TFs) is one such mechanism which accredits entities with experience, platform independence, and skill in identifying and reporting illegal content. 
With the DSA's TF role being roughly one year old,
we interviewed representatives of seven such TF organizations to learn about their experiences of becoming a TF and how it impacts their interactions with online platforms and with individual users. 
We additionally ran a workshop involving TF representatives, primarily as it was requested by TFs themselves, who collectively wanted to share experiences of their new role and learn from each other rather than be isolated. 
Notably, we found that accreditation as a TF can be cumbersome, that resources for TFs remain the same despite an increasing workload, and that platforms priorities often diverge from TFs.
We conclude with recommendations for future research into understanding user representation within the DSA and the need for standardization measures tailored to the needs and resource constraints of TFs.
\end{abstract}




\maketitle

\section{Introduction}
Online platforms are a central component of today's interaction between individuals, whether it is to communicate~\cite{tang_impact_2021}, find information~\cite{kim_individual_2014}, shop~\cite{varghese_impact_2021}, or just be entertained~\cite{xia_navigating_2025}. 
However, these online spaces pose different risks to users – one of which is illegal content online. 
To have a regulatory answer to this and many other risks
the EU has created the Digital Services Act (DSA)~\cite{noauthor_regulation_2022}. 

This novel regulation forms obligations for transparency, assessment, reporting, and design for online platforms like Instagram, TikTok, X, or Pornhub, as well as for search engines (See Art 33 DSA~\cite{european_commission_overview_nodate}). Other countries are also pursuing comparable approaches to regulating online platforms and their processes for responding to harmful content, like the UK~\cite{law_effective_2024}, India~\cite{noauthor_intermediary_2023}, or i.e. in the US specific rules in Florida~\cite{noauthor_senate_2021} and Texas~\cite{noauthor_texas_2025}.

The DSA includes transparency obligations (Arts.~15,~24,~42,~17,~39 DSA), systemic risk assessment and mitigation (Arts.~34-35 DSA), independent audit structures (Art.~37 DSA), or new user facing redress mechanisms (Arts.~20-21). The DSA also creates Trusted Flaggers (TF) as a role to patrol the online space and identify illegal content online (Art.~3(h) DSA) that has eluded the prior moderation process (Art.~22 DSA). A TF is accredited status by the DSA's national regulators.\footnote{Digital Service Coordinators according to Art.~22(2)DSA.}
Online platforms must prioritize TF reporting of illegal content over normal user flagging (Rec.~61).

It remains unclear what the challenges are for organizations and individuals that pursue TF accreditation, and how their relationships with regulated platforms develop.
This is especially the case given that the TF role as a mechanism is only a matter of months old at the time of writing~\cite{european_commission_trusted_nodate}. 
Expectations of TF organizations within the DSA are not yet well-understood and pose a research gap~\cite{sekwenz2025unfair} due to the novelty and legal complexity of the work.

We aim to understand how accredited TFs perceive their role in reporting content and engaging with the platforms hosting content, as a new experience beyond the mechanisms previously available to them, such as Google's Priority Flagger program~\cite{google_about_nodate}. We also contrast the capabilities and requirements of the TF role with the perception of regulatory activities within the role, and the extent to which the role enables them to achieve their goals in monitoring illegal content online. With this, our overarching Research Question (RQ) asks: 
How do Trusted Flaggers enact the Article 22 DSA mechanism in practice, and what frictions and accountability implications arise from this for user representation?


Our Research Questions (RQs) are:
\textbf{RQ1:} What have been Trusted Flaggers’ (TFs) experiences in becoming and acting as a TF, particularly in comparison to their activities before obtaining TF accreditation? 
\textbf{RQ2:} What gaps emerge between the capabilities the DSA assigns to Trusted Flaggers (TFs) and the resources, tooling, procedural support, and public visibility? 
\textbf{RQ3:} How do Trusted Flaggers’ (TF) experiences reveal differences in how platforms operationalize and prioritize DSA content reporting obligations, what are key challenges for TFs, and how are these practices reflected in DSA transparency and accountability artifacts (e.g., transparency reports and Statements of Reasons)? 

The paper is structured as follows: Section~\ref{background} outlines the Trusted Flagger mechanism under the DSA and its relationship to notice-and-action and transparency obligations. Section~\ref{related_work} reviews related work. Section~\ref{method} describes our methods. Section~\ref{results} reports the findings. Section~\ref{discussion} discusses implications and recommendations for DSA implementation and accountability. Section~\ref{conclusion} concludes the paper.

\section{Background}
\label{background}
Article~22 DSA introduces a ``fast-lane'' for illegal content (Art.~3(h) DSA) notices (Art.~16 DSA) submitted through content reporting interfaces by a Trusted Flagger (TF) (Art.~22 DSA) \cite{rosati_dsas_2024}. TFs are positioned as an institutional response to illegal content that harms users 
(like CSAM content or disinformation). 
Through their mandate-bound flagging activity, TFs operationalize specialized expertise in the identification and substantiation of presumed illegal content, in a way which overcomes the challenges in reporting content which non-experts would face, in terms of collecting evidence and understanding the law 
\cite{sekwenz2025unfair}.

TF status (accreditation) is granted by national regulatory authorities of the DSA – the  Digital Services Coordinators (DSCs) (Art.~22 (4),~(7-8),~49-51) – for a designated area of expertise (within a specific mandate of illegal content). This status is conditioned on demonstrable competence, independence from platforms, and diligent, accurate, and objective flagging (Art.~22(2)). The status as a TF can be suspended and revoked by DSCs in cases where notices are imprecise, inaccurate, or inadequately substantiated (Arts.~16,~20,~22(6)-(7)) \cite{appelman__2022}. The DSA also embeds transparency and accountability requirements through annual public reporting (TF reports) and regulator-facing disclosures (Art.~22(3)). The DSA links TF reporting activity to platform transparency reporting (Arts.~15(1)(b)) and statements of reasons (Art.~17(3)(b),~22(3)-(5)), as well as in systemic risk mitigation measures (Art.~35(1)(g) \cite{sekwenz_digital_2025, kaushal_automated_2024}. 


For the FAccT research community, TFs offer a concrete case of how regulation reshapes content moderation practice. TF designation introduces procedural prioritization (Rec.~61–62) that structures the timing of platform decisions, an aspect with normative consequences for affected parties beyond efficiency considerations \cite{susser_decision_2022}. TFs also change the evidentiary and substantive quality of notices by adding external expert reporting. Finally, TFs foreground questions of representation and accountability (Art.~86), because particular organizations are empowered to escalate certain harms and illegality claims within platform processes. These interactions can produce new dynamics for the actors involved and point to open questions about how best practices should be developed and institutionalized.

\section{Related Work}
\label{related_work}
\paragraph{Flagging as content moderation measure}
Content reporting under the DSA (or, flagging \cite{crawford_what_2016}) is a form to bring problematic content to the attention of platforms through using UIs \cite{shim_incorporating_2024}, either by users (as a community-based approach \cite{kou_community_2024}) or by experts like TFs (which has the potential to changes effectiveness \cite{garrett_flagging_2019} or user beliefs \cite{lanius_use_2021}).

Recent work operationalizes arbitrariness in algorithmic moderation as measurable inconsistency in decision outcomes, linking it to procedural justice and non-discrimination concerns \cite{gomez_algorithmic_2024}. By introducing the new role of TFs the procedural order of content moderation on platforms changes too. Under the DSA platforms should enforce their content moderation decision however in a ``non-arbitrary and non-discriminatory manner" to enable diligent processes.
It is still unclear how to test such semi-automated processes for compliance \cite{terzis_law_2024}, and links back to general questions of assurance in auditing algorithmically supported processes \cite{lam_framework_2024}. Evidence from crowd-sourced moderation suggests that reviewer composition shapes political bias and moderation outcomes; TF's role can be discussed as a contrasting ``expertise-credentialed" structures that may mitigate or reconfigure these biases in the future \cite{thebault-spieker_diverse_2023}.
New regulation such as the UK Online Safety Act (Sec.~20) \cite{noauthor_online_2023} or the DSA (Art.~16) creates new obligations for how such reporting interfaces must be designed \cite{sekwenz2025unfair}. 

Flagging can be used to open up the content moderation queue (and result in moderation decisions like deletion) \cite{naab_flagging_2018}, or soft moderation measures like the application of warning labels \cite{barman_discerning_2024, jamieson_flagging_2025}.
Flagging as a form of counter measure \cite{chua2019identifying}, or `trust and safety' \cite{moran_end_2025, disalvo_social_2022} to mitigate harmful content online \cite{arora_detecting_2023}
can be seen as the human counter part to algorithmic content moderation \cite{gorwa_algorithmic_2020, sarwar_neighborhood_2022, devadas_experimental_2024, perez_analyzing_2025}.
Especially differentiating `accurately' \cite{wei_operationalizing_2025} between legal and illegal content is challenging \cite{yar_failure_2018, wagner_mapping_2024}, even for widely accepted topics like CSAM content national definitions differ \cite{kokolaki_investigating_2020}. However, under which legal reason such reports are filed matters for transparency and measurement of risks \cite{wagner_regulating_2020, holznagel_follow_2024}.
content moderation can be perceived as fairer by users \cite{cai_content_2024, ma_im_2022, teblunthuis_effects_2021} as a form of procedural justice mechanism \cite{katsaros_online_2024}. The design and transparency of such mechanisms, on the one hand, can influence flagging behavior \cite{zhang_cleaning_2024}.
On the other hand, content moderation measures can be misused to silence specific groups \cite{are_assemblages_2024, lyu_i_2024}, i.e. content reporting can be used to silence marginalized or at risk groups \cite{are_flagging_2025}, be perceived differently by various user groups like young users \cite{baumler_towards_2025} or users with different accessibility needs \cite{hosamane_i_2025}, or can create privacy concerns for users \cite{li_obfuscation_2022}. 

\paragraph{A Legal Lens on Trusted Flaggers}
Legal scholarship has argued that TF arrangements function as \emph{privileged} third-party access to platform notice-and-action, and that their legitimacy is contested because they promise expertise while raising concerns about opacity, accountability, and over-removal. Appelman \& Leerssen \cite{appelman__2022} conceptualize trusted flagging as a heterogeneous governance practice that is varying by legal construction, degree of privilege, and point of intervention in moderation, and map competing narratives that frame TFs as (i) a source of expertise and inclusion, (ii) an unaccountable co-opting of public/private power, or (iii) a performative signal of inclusion rather than a structural shift in control.
Van de Kerkhof’s constitutional analysis \cite{van_de_kerkhof_constitutional_2024} highlights how formalizing TF roles can intensify freedom-of-expression and rule-of-law concerns, especially where public authorities act as TFs and risk ``censorship-by-proxy'' through platform enforcement. He emphasizes the need for stronger legality and proportionality safeguards and clearer constraints on state-originating flagging practices \cite{van_de_kerkhof_constitutional_2024}.
Prior work has examined TF arrangements as privileged, expertise-based reporting channels that platforms may prioritize, often through bilateral and opaque partnerships. Van de Kerkhof analyzes how Article~22 DSA formalizes this model while foregrounding persistent tensions around over-removal, transparency, and fundamental-rights \cite{van_de_kerkhof_article_2025}. The author further highlights implementation risks that matter for socio-technical studies of notice-and-action (Art.~16 DSA), including the contestability of certification criteria (e.g., ``objectivity''), cross-border legality frictions (e.g. the definition of `illegal'' content under Art.~3(h)), and potential gaps in user-facing transparency when TF reporting is operationalized through platform workflows \cite{van_de_kerkhof_article_2025}. Finally, the paper proposes harmonization and standardization measures (e.g., under Art.~22(8) and Art.~44(1)(c)) and discusses resourcing options for TFs \cite{van_de_kerkhof_article_2025}.

\section{Method}
\label{method}
\begin{figure*}[ht]
    \centering
\includegraphics[scale=0.33]{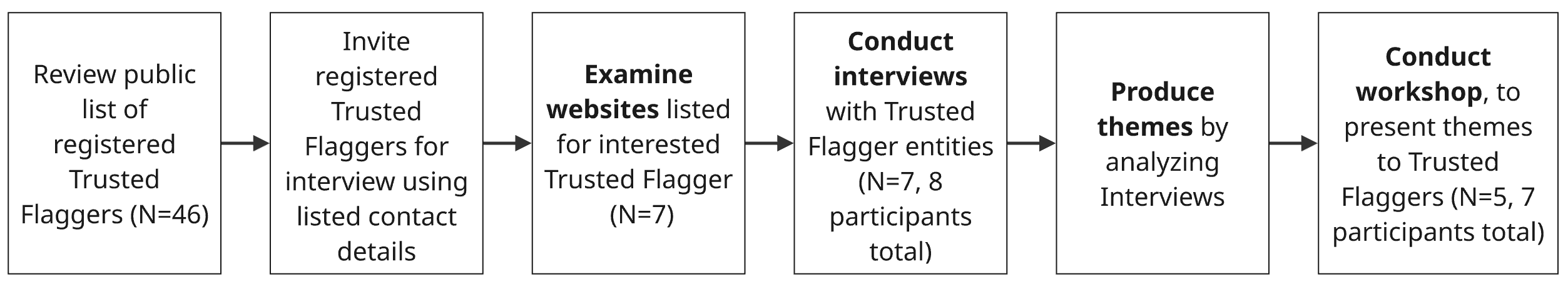}
    \caption{Steps in our engagement with Trusted Flagger entities.}
    \label{fig:method}
    \Description{Steps in our engagement with Trusted Flagger entities.}
\end{figure*}

Here we describe our multi-stage approach to understanding Trusted Flagger activities. The flow of research activities is illustrated in Figure \ref{fig:method}. As this study involved human-subjects, we included details about research ethics review later in the paper, in Section~\ref{sec:ethics} (Ethical Considerations Statement).

\subsection{Semi-structured interviews}
\label{semi_structured_interviews}
We conducted seven interviews with eight participants (one interview had two participants). The interviews were semi-structured. 
We began with questions that we wanted to ask, but we remained open to hearing what participants wanted to speak about and emphasize from their own perspective.
The full set of interview questions can be found in Appendix \ref{Interview_Guidelines}. Table \ref{tab:participants_scope_workshop} illustrates the position of participants as well as the organizations' scope. Additionally, it shows workshop attendance \ref{sec:meth-workshop}. During interviews, participants were shown screenshots of the Statement of Reasons database (covering all Trusted Flagger reports across platforms from 25 September 2023 to 5 November 2025) to elicit views on DSA accountability artifacts (see Figures~\ref{fig:SOR_1} and ~\ref{fig:SOR_2}; \cite{european_commission_dashboard_2025}).
Each interview lasted approximately one hour. Interviews were conducted using Microsoft Teams and its automatic transcription function.


\begin{table}[ht]
  \centering
  \footnotesize
  \begin{tabular}{@{}p{0.9cm} p{4.0cm} p{3.6cm} p{1.6cm}@{}}
    \toprule
    \textbf{Part.} & \textbf{Role / Position} & \textbf{Organization Scope} & \textbf{Workshop} \\
    \midrule
    P1 & Lead TF team & Online information integrity and public-interest harms. & Yes \\
    P2 & \textbf{P2-1:} Legal counsel;\newline
         \textbf{P2-2:} Leads communication/cooperation with platforms
       & Consumer- and product-related illegality. 
       & \textbf{P2-1:} Yes;\newline \textbf{P2-2:} No \\
    P3 & Scientific researcher & Identity- and belief-related harms and illegality. & No \\
    P4 & Legal advisor and content analyst & Interpersonal harm prevention. & Yes \\
    P5 & Cybercrime specialist & Online harm-related activity. & Yes \\
    P6 & Hotline manager; lead analyst team & Online safety reporting. & No \\
    P7 & Team lead TF & Consumer-related illegality. & Yes \\
    \bottomrule
  \end{tabular}
    \caption{Overview of interview participants, including role/position, organizational scope, and workshop attendance.}
  \label{tab:participants_scope_workshop}
\end{table}

\paragraph{Interview recruitment and participants}
\label{interview_recruitment_and_participants}

As in Figure \ref{fig:method}, we consulted the public list of accredited Trusted Flaggers (TFs), which includes contact information \cite{european_commission_trusted_nodate}. At the time of the recruitment there were 46 registered Trusted Flaggers, and at the time of writing there were a further 10 TFs accredited.

We sent an interview invitation to the listed contact email address of each TF in the list on the 30th of October 2025 and waited for responses of TFs until the 26th of November. Representatives of seven TFs responded positively, which then led to interviews. We had two other positive responses, though one was a fixed written response and the other was received after the workshop had already been conducted. Our response rate for recruitment was then 15\% of accredited TFs at the time of the research -- the community of organizations to recruit from is currently very small but growing, and our response rate was comparatively high.

For organization `P2', two representatives joined the same interview call separately (P2); we identify these as P2-1 and P2-2 (as in Table \ref{tab:participants_scope_workshop}). As the TF directory is a public list, we knowingly define Organization Scope in a manner that is high-level but is indicative of participant roles. 

\paragraph{Participating organizations -- website analysis}
\label{website_method}
To answer RQ2, we also conducted an analysis of the participants' organization websites.
The website analysis allowed us to investigate how organizations communicate their Trusted Flagger status to users and how Trusted Flaggers expect to interact with users.
We first located these websites using the public list of accredited TFs. 
One author then proposed a set of categories upon which to interpret how components of the websites related to the research questions. 
These categories were then discussed and refined among all authors.
The same author then noted websites information in those categories (such as languages used, 
and user contacting process
), so that we could identify how the various TF websites communicate what they do, who they support, and whether they mention that they are a TF.  
Insights gained from the website analysis were carried into the interviews.

\subsection{Interview analysis}
\label{sec:analysis}

Interview transcripts were analyzed using reflexive thematic analysis (RTA)~\cite{braun2021one}.
This enabled us to identify the capacity, processes, and perceptions of Trusted Flaggers when they engage with platforms, regulators, and users.
All authors began by familiarizing themselves with the data by reviewing the transcripts.
Two authors independently conducted open coding on the first three transcripts.
All authors then met to cluster codes around certain themes before continuing open coding the remaining transcripts. 
The first coder had 35 codes and the second coder had 27 codes, yielding a total of 62 codes between them.
We did not measure inter-rater reliability given the iterative and reflexive nature of our analysis~\cite{mcdonald2019reliability, braun2019reflecting}. However, there was considerable overlap in the two sets of codes, supporting further discussion.
Instead, the codebooks of the two coder-authors were used as reference artifacts during discussion, given the reflexive nature of the analysis and that they have their own research backgrounds and experience relative to the study.
While the two coders were aligned on the broad themes developed, the first coder focused on identifying the perceptions of TFs (especially through a lens of fairness) while the second coder aimed to identify the processes and interactions which TFs engage in.
Both coders had experience conducting and analyzing interview-based studies prior to this study. 
All authors met or communicated regularly (at least every few days) to discuss the coding process.

Six themes emerged from the reflexive interview analysis: \textbf{Platform frictions:} Platforms obstruct or fragment Trusted Flagger work; \textbf{Capacity constraints:} Capacity, resourcing, and training constraints shape what Trusted Flaggers can do; \textbf{Coordination challenges:} Regulatory coordination is uneven and (sometimes) difficult; \textbf{Platform incentives:} Platforms' incentives and response patterns shape outcomes; \textbf{Operational ambiguity:} Illegality is operationally hard to pin down in practice, and; \textbf{Accountability artifacts:} Transparency and accountability artifacts can misrepresent Trusted Flagger labor and role. These themes are represented in the Results section as Sections \ref{sec:res-frictions} to \ref{sec:res-account}). One purpose of the subsequent workshop (Section~\ref{sec:meth-workshop}) was to validate the six themes with participants; no participant disagreed with the presence of any of the themes.


\subsection{Validation workshop}
\label{sec:meth-workshop}

We planned to conduct a workshop participating organizations, contingent on there being sufficient take-up of interviews and interest in a workshop. In reality, several participating interviewees explicitly requested a chance to meet other Trusted Flagger organizations and were curious about our broader results. 
Seven participants joined the online workshop, from five of the  organizations represented in interviews (Organization 1-2, 4-5, and 7, see Table \ref{tab:participants_scope_workshop}). More participants from organization P4 attended the workshop, as they were particularly interested in meeting other TF organizations. 

The workshop was structured around an overview of the themes which emerged from the interview analysis, as described in Section \ref{sec:analysis}. We began the workshop by stepping through a set of slides; each theme had one slide, which included 2-3 additional statements relating to each theme, for which we sought to understand (dis-)agreement, and which otherwise served as prompts for discussion (to validate the themes). Early presentation of overview themes was also an effort to make attendees feel comfortable voicing opinions on each theme afterwards. To this end, we encouraged attendees to speak up at any time with comments. After the presentation of themes, we encouraged open discussion, where we intended the workshop to also serve as a co-creation opportunity for Trusted Flaggers. To lower the barrier to participation, we supported both spoken interventions and written input during the workshop. The workshop lasted 90 minutes, as scheduled. 
Two of the authors were present, with one presenting the results, and the other present to monitor discussions. 
We detail broad outcomes of this discussion in Section \ref{sec:res-workshop}. 


\subsection{Limitations}
\label{limitations}

We recruited Trusted Flagger practitioners on a best-effort basis - we contacted the listed contact address in the directory of TFs, but only made one attempt, with an initial email invite. We did not send reminders. Between sending the initial invites and the planning of the workshop, there was approximately four weeks, allowing TFs to respond. Only one positive response arrived late. The interview analysis did not result in any new codes after the first five interviews, from either coder.  

Many participants reported only recently gaining Trusted Flagger status. These are early days for the DSA, so many activities have yet to become routine. However, our analysis already identified early challenges, where not all of these challenges will necessarily be resolved over time, but may require additional attention (e.g., platform opacity). 
We did not speak to the platforms or regulators themselves. 
We identified several shared dynamics of interaction from the interviews, between TFs and these entities, indicating that they are not isolated experiences. 

\section{Results}
\label{results}
Here we detail the results of our interview and workshop analysis. Sections \ref{sec:res-activities} and \ref{sec:res-users} provide foundational context for our participant cohort and their activities. Sections \ref{sec:res-frictions} to \ref{sec:res-account} describe the themes identified from the reflexive interview analysis \ref{sec:analysis}. Closing the results section, Section \ref{sec:res-workshop} describes the outcomes of the online workshop we conducted with representatives of several participant organizations.

\subsection{Trusted Flagger activities}
\label{sec:res-activities}

The process of becoming a Trusted Flagger (TF) could take several months (P1, P4). P7 attributed their 6–7 month process of accreditation work to there being a lot of questions and paperwork which ``took for us more than 200 hours of multiple people working on it.'' The process requires financial and organizational data, as well as proof of language and legal capacity, to be submitted to the national regulatory authority (the DSCs).


TFs must define their mandate, i.e., their area(s) of expertise and the categories of illegal content they are authorized to report. They must also evidence their ``capability to identify'' content as stated by P7. According to  Art.~22(1)(a-c) TFs must provide evidence of competence, impartiality, and objectivity within their area of expertise. The TF role therefore demands additional `capabilities' (P1). TFs follow a strict scope where they do not step outside their mandate even when contacted about content beyond it (P4), especially as ``there's already so much and we cannot do it all'' (P5). P6 added that demonstrating platform independence requires evidence that their organization did not receive platform funding, or only negligible amounts. 

Motivations to become a TF included the status the role brings and its `highly relevant' value for their activities (P2-1), as well as interest in how the DSA unfolds – P7 spoke of their curiosity, and P5 described it as ``a first kind of pilot phase.'' 
P5 saw involvement in the DSA as an opportunity to influence regulation in ways that better reflect their organization’s cause.

\textbf{TF activities as a new routine:}
\label{TF_activities_as_a_new_routine}
Participants described TF entities as supporting both enforcement and recovery, for example by helping owners of `credible' accounts (P1) that were taken down to restore them, addressing situations where reporting or take-down mechanisms were used against non-harmful entities. P1 also noted that the DSA’s scope is ``much more narrower'' than TFs’ day-to-day practice, because it focuses on illegal content, whereas in practice, TF work often extends to examining harmful or ``highly problematic'' content that does not meet the illegality threshold. 

P1 described routine activities of ``finding, vetting, checking, and documenting content, preserve[ing] the evidence,'' and writing the report. This was mirrored by P2-2, who emphasized that reporting centers on ``accuracy, compliance with all legal requirements and platform policies''\footnote{Platform policies refer to platforms' Terms and Conditions (TaC) according to Art.~3(u) as "[...]] all clauses, irrespective of their name or form, which govern the contractual relationship between the provider of intermediary services and the recipients of the service." and further regulated in Art.~14 DSA.}; 
Documentation work would also capture ``URLs, [...] infringement identifiers'' and other supporting information (P2-2). P4 explained that they use tables, combined with a series of steps, to identify the relevant contact-point addresses within specific organizations and then route reports accordingly (
in some cases also to law enforcement).

After submitting reports, TFs must monitor how platforms respond, which adds a further task that scales with the number of outstanding reports. P2-2 also noted that reports may be `rejected' or that platforms may request ``additional information.''
Since TFs must substantiate an annual report about their activities, they must also evidence their work through, e.g., ``screenshots of the violation[s],'' as outlined by P7. Different approaches to this documentation process appeared to develop at each TF.

\textbf{Costs of TF activities:}
\label{Costs_of_TF_activities}
Where there are regular, routine activities of TF entities, participants described accompanying costs. The experiences of participants resembled the longer end-to-end explanation provided by P1:

\begin{quote}
\textit{``you need to have the monitoring, you need to train the people. They need to find the specific violations, tailor all the documentations, collect the evidence, store the evidence. [...] then write report [...] All of this evidence put in the document, then register in the registrar where we need to measure when it's reported and how much time does it take to take it down and then split the report to 20, 30 or 60 reports and then submit it, then track if you received confirmation e-mail and what is the timestamp of that confirmation e-mail. So you could put it in the Excel sheet so you could measure when it will be taken down. Then you will wait for the response and then when the response comes you will need to copy paste and check when that report was taken down so you could later write down the final report.''} (P1)
\end{quote} 

These activities take effort, but already in these early stages of DSA roll-out, are chipping away at the resolve of TF members, who -- where there are not additional resources to cover the costs -- are acting somewhat as volunteers. P1 points to the hurdles of the reporting process that seem to be ``designed to make your work miserable'' and that every platform handles reporting interfaces and processes differently, which collectively impacts the quality of TF work in that ``we [...] need to do [a] crazy amount of things, but you just are so under resourced that you just are not capable to do [that well].'' 
P3 noted that ``there is no specific budget allocated for'' DSA-related activities in their organization, such that it is ``basically just an extra task that come on top of all of the other things'' they do in their role.
P5 noted that their organization had been exploring ways to have platforms themselves allocate more resources to moderation of content (rather than it being them who would have to do so).


\subsection{Trusted Flagger activities and users}
\label{sec:res-users}

Many of our participant Trusted Flagger (TF) entities did not directly interact with members of the public in a reactive way. Rather, the TFs had a cause that drove their activities, which was seen as benefiting society, for instance P1, ``Our job is to expose [...] illegal things and inform users.'' 

However, some TFs regard user reports as their starting point (such as P4, which would typically ``receive [a] report from the public'' or trusted organizations like NGOs). User reporting was more essential for P6 and P7, whose organizations were already dedicated to directly interacting with users before the DSA. These reports sometimes came from users who were ``aware of our role as Trusted Flaggers'' and included ``thorough documentation and clear legal reasoning'' (P7).
However, the TF status is only one tool for these organizations as ``90\% of the reports we handle come from the online form'' (P6). 

P4 noted that they would sometimes be contacted by ``someone I guess that tried everything they could before,'' who they would then assist. 
Others, such as P1, do not routinely interact with individual users or citizens and do not see reporting for users as a key TF task.
TFs such as P3 and P5 are somewhere in-between, being contacted about content but responding on a best-effort basis, informed also by the amount of detail provided by the individual user (with it being ``a matter of capacity'', P5). This adds shape to the developments noted in prior work \cite{sekwenz2025unfair}, where the participants of the Sekwenz et al. study were not sure how TFs and users should be aligning with each other.

\textbf{(Lack of clarity on) Outward-facing visibility and role:}
When discussing TF visibility, such as the necessary online presence and website for a TF, it would appear to be something regulators and TFs still need to clarify (P1). 
P6 noted that ``[TF status] is a very precious thing, but I don't think it was very well communicated to the public." Beyond any lack of clarity, another view among TFs was that 
it did not appear to change the nature of their interactions with users. P3 did though note that as a TF, they will be publishing public reports.

\textbf{Responding to users who challenge reports:}
During the interviews, not all mention of users was about those individuals who shared the cause of the TF. In some cases, there were platform users who were the very same actors that TFs were acting against, i.e., those parties considered to be proliferating harmful material. This included actors seen as using DSA mechanisms to have their reported accounts reinstated, as noted by P1, where ``we again spend our time,'' as (in particular) `big' accounts were ``Reappearing after [...] we receive information that [...] they were taken down. So some of these big pages were taken down already three times and now the four times again they are being restored.'' 


\textbf{Visibility via websites:}
Looking at how TFs communicate with users, we found that there was no standard way organizations presented their TF status. Organizations used blog posts, press releases, and social media posts to communicate their TF status. The websites were also particularly helpful in understanding how the organizations' working areas compared to their stated area of expertise under the DSA--with three organizations selecting more areas of expertise than their websites communicated.

Some organizations included instructions on how users could contact them. 
The websites were largely accessible by people who spoke different languages as the participants' websites covered a total of 10 languages.
All organizations were transparent of their funding sources, whether they came from government sources, donations, corporate memberships, or even the platforms themselves.

\subsection{Platform frictions}
\label{sec:res-frictions}

Platforms were seen as perpetually having content violations which needed effort to report and address, such that they would 
``communicate with the platforms on a daily basis'' (P2-2).  

Views on the platforms' approach to DSA implementation were mixed, especially when it came to the differences in reporting pathways both within and between platforms. This was either seen as the reality, but also potentially intentional, that ``[platforms] 
you know, they're doing everything to make this work complicated [because] we take out the revenue'' (P1).
This can result in a ``frenemy situation'' where TFs offer help and support but ``they don't treat us as friends.'' 
(P1). 
P3 highlighted an additional dynamic that tilted in favor of platforms, where ``it also feels like it's a way to externalize the responsibility for hateful or illegal content that should belong to these social media platforms that are making huge profits from what users post.''

P7 found that with various reporting channels now existing, it ``makes it difficult for us to especially choose the right reason we're reporting to get on the right path for reporting.''
More practically, some platforms limited the amount of reports TFs could make, with P5 only being able to ``report 10 URLs max at one time" after ``they send you a link that gives you access for one hour [...] and if you were too slow, you have to ask again."

P5 noted that some `smaller' platforms are outside the Trusted Flagger provision, such that there not only is ``no harmonized flagging system'', but that if a website has no mention of the DSA, it can be ``quite hard to find how to contact them.'' 
That TF experiences with smaller platforms can differ from those with Very Large Online Platforms (VLOPs) resonates with themes identified in research that precedes the passing of the DSA \cite{laux2021taming}, where Laux et al. argue that the larger platforms will concentrate demand for audit activity. 
P2 and P3 noted roundabout discussions with platforms, that sometimes platforms re-request information and afterwards ``ghosted'' TFs (P3).
P3 also speculated that the responses they receive from TF reports may be automated.

\subsection{Capacity constraints}
\label{Capacity_constraints}
Participants were clear that they cannot act beyond the mandate they declare in their Trusted Flagger registration. The mandate collided with resourcing especially when considering irregular contact from individual users or other organizations, as ``we are thinking how to support them. But this is again kind of problematic, you know, because it's to support them, it's cost for us'' (P1).

\textbf{Tooling:}
Many participants reported having developed, crafted, or obtained dedicated monitoring tools as a way to manage their work and streamline their efforts (e.g., P1, P4), This included AI-driven tools which automatically scan platforms for `infringing content' (P2-1) and ``automatically generate a list of potentially illegal content'' (P6). These tools may be specific to their cause, raising questions as to how transferable those tools may be for others to learn from (as became a theme during the later workshop, Section \ref{sec:res-workshop}). 

It was seen as necessary to bolster monitoring tools with subsequent data processing and classification capabilities. Crucially, this suite of capabilities then needed review by humans (P1, P5), or needed human involvement as part of engagement with other stakeholders (P2-1). 

\textbf{Resourcing TF activities:}
Several participants gave shape to the limitations within which Trusted Flagger (TF) activity happens. For instance, 
``we don't have dedicated people who only work on this [...] Like how can we dedicate people if we don't have resources on that?''
(P1). As such, many employees would be involved in trusted flagging ``on a rolling base, whoever has time" (P7) rather than being the task of a single employee. 
Similarly, P3 emphasized that TF activities were in addition to a range of other duties and roles they otherwise had within their organization, without additional budget. P5 framed human resources and expertise as ``our biggest asset.''

P1 emphasized the `vulnerability' of facing platforms in court cases. 
This was described in the context of providing evidence for DSA infringements where platforms in court  
``need to discredit our evidence'', such that 
``we're making ourselves vulnerable, working for free" (P1).
The scrutiny needed to protect against such scenarios is, in the case of P7, a mandatory ``four eyes principle" for content evaluation. 


Some TF entities may have membership payments which cover their costs (perhaps as a non-profit organization), whereas others may be supported by volunteer efforts (much like a charitable organization). However, TF organizations struggle to fund flagging activities in addition to their existing activities.

\subsection{Coordination challenges}
\label{Coordination_challenges}
Becoming a Trusted Flagger (TF) was seen as a way to smooth the path, 
as necessary to ``flag non-cooperative platforms'' to a DSC or the European Commission (P2-1). 
P1 on the other hand saw it as ``we're doing institutions work" and therefore not only supporting content moderation processes of platforms, but also contributing to DSA enforcement, needing to coordinate across 
DSCs, the platforms, and the local authorities. 
This coordination with stakeholders was further highlighted by other participants, referring to national law enforcement and judicial powers (e.g., P1, P4, P5). 
Similarly, P6 referred to their trusted flagging activities as happening ``almost parallel" to law enforcement.


P2-1 noted being in regular communication with the DSC and Commission, regarding clarification of guidelines and procedures; this was echoed by P4. P4 cautioned that 
there can be cases where a platform is ``not compliant or does not want to cooperate'', where P7 noted similar experiences and a need to approach the DSC, after they had 
``filed a number of complaints of Trusted Flagger cases with [a platform]'' and they were repeatedly declined.
P4 discussed that (aspiring) TFs are in need of templates for notifications to platforms, as well as details on annual TF reporting duties. We also saw that -- perhaps due to the DSA still finding shape -- TFs would also ``write some recommendations'' (P1) to be taken, ultimately, to the Commission. 

\subsection{Platform incentives}
\label{Platform_incentives}
Many participants noted that the effort required to report was compounded by platforms, big and small, having different reporting pathways. As remarked by P1, ``with every platform, the process, how you submit the reports is different. There is no single form," but rather every platform designs reporting and TF mechanisms differently. 

Some platforms reached out to TFs to establish a reporting channel, though this did not always happen smoothly, 
where there were anecdotes of how a platform may require a TF ``to create an account, [yet] it asks for a phone number, but a personal phone number. But obviously as an organization [...] we cannot use our private ones because then we already have a [private] account" (P5). 
The cumbersome nature of reporting did serve to remind TFs that ``it's a zoo and it's difficult for us" (P7) and ``this field is still very, very, very new'' (P1). 

In contrast, P2-1 and P2-2 noted that platform policies may fill in gaps in EU regulation, where those policies provide additional mechanisms useful for the cause of the organization outside of its TF status (such as `defamatory comments'). However, they also stressed the need to ``have a more uniform way of implementing the trusted flagger status'' (P2-1).

It was seen that the move to structured DSA reporting has resulted in more cumbersome, slower reporting pathways. This has created a situation where 
``some platforms apply unnecessary requirements to take down notices'' (P2-1). As noted by P1, ``with Meta we can only report 20 URLs at a time. 
So that means that we will need to instead of sending one report. [...] They are taken down bit by bit, 20 by 20,'' whereas in the past, ``we were submitting by emails and they were kind of reacting.'' 

\subsection{Operational ambiguity}
\label{Operational_ambiguity}
The type of content being reported often made it difficult for participants to determine legality.
P2-1 noted experiencing ``a very high success rate'' with their reports, 
but that ``IP infringing content is a little more black and white...rather than maybe some gray or political content or something like that.''
On the other hand, P6 found that platforms were slow to delete ``accounts that are specialized into linking [illegal] content that's not on the platform which is external to the platform."
P7 explained that platform features pose different challenges, such as Reals on Instagram that `vanish' and can not easily be reported or evidenced for TF activity.
P5 often struggled to determine the legality of content where external documentation was required but only alluded to in the content. 

\subsection{Accountability artifacts}
\label{sec:res-account}

When presented with the reporting data captured by the Statement of Reason (SOR) database including all TF reports Participants variously had difficulty in relating the platform reporting data to their own reporting activities, where ``It does not reflect reality'' (P1), and ``it just feels not very transparent'' (P5). 
Compounding this, as P1 noted: ``Every platform has different community standards" with overlaps of about 80-90\%, these however are formulated in a always different and ``very unfriendly way" demanding time of TFs.


Several participants noted difficulties in relating statement of reasons data with their own activity when presented with the screenshots illustrating TF activity (See Figures \ref{fig:SOR_1} and \ref{fig:SOR_2}), for instance P3, commenting that ``maybe even under 10 [reports] were in the database while we had done 10s of [reports],'' leading them to wonder if the database is ``not updated properly''. 

Participants were left to come up with their own justifications for why reports did not show up in the public database.
Participants hypothesized that the terms of service was a wide umbrella (though not the correct one) which covered illegal content implicitly (P6), but also significantly plays down the TF activity – who \textbf{only} report \emph{illegal} content through reporting mechanisms.
Meanwhile P7 argued, ``the platform steers you into what is favorable for them...if we were certain we flagged illegal content under specific laws, the platforms came back to us saying, `thank you,' but this doesn't violate our terms of services."

\subsection{Workshop outcomes}
\label{sec:res-workshop}
The workshop primarily functioned as (i) a check of our interview-derived themes among participants (See Section \ref{results}) and (ii) a peer-to-peer exchange, where participants had an opportunity to compare platform-specific reporting workflows, practices, and interactions.
During the workshop discussion, participants corroborated the themes identified in our analysis. This led to discussion of practical day-to-day experiences, as below.

Participants collectively stressed that platform implementations can actively shape what becomes visible as TF action. For example, in the discussion of Meta’s TF portal constraints (e.g., URL limits per submission, see Section \ref{TF_activities_as_a_new_routine}, \ref{sec:res-frictions}, and \ref{Platform_incentives}), participants compared their technical fixes, such as 
adding additional URLs in free-text fields. There were then concerns that data registered in non-standard ways 
may not survive the journey through to platform confirmations or downstream transparency and accountability artifacts such as the SOR database (
see Section \ref{sec:res-account}). 

The workshop also underscored that measuring ``prioritization'' as a main TF quality compared to other user reports is currently methodologically under-specified: participants debated whether take-down time should be computed from the moment of submission versus platform acknowledgment of `receiving' TF reports, how to evidence submission timestamps, and how missing confirmations and partial/batched processing distort measurement. Some participants would illustrate their view by discussing their own tooling for reporting 
(see Section \ref{Capacity_constraints}).

Another core topic of discussion was cross-member-state variability in accreditation and oversight practice (e.g., differing requirements for demonstrable legal expertise). There was a call for more systematic on-boarding by regulators -- such as maintaining contact points and reporting-channel information for platforms -- to reduce duplication of effort when TFs are newly designated (see Sections \ref{sec:res-activities} and \ref{Coordination_challenges}). 
Similarly, 
there was discussion of practical coordination among TFs, for example by sharing templates, metrics, and ``lessons learned'' 
(see Sections \ref{sec:res-frictions} and \ref{Platform_incentives}). 

\section{Discussion}
\label{discussion}

Here we discuss how our engagement with Trusted Flaggers (TFs) links to outstanding challenges in both the emergence of DSA infrastructure and user support more generally. 

\paragraph{The opaque nature of large platforms and providers}

A common narrative among our participants was the large scale of content that they were seeking to manage. 
If the reporting pathways on one platform are cumbersome, it absorbs attention because Trusted Flaggers have limited resources -- each platform then `benefits' mutually when each other individual platform is difficult to report through, in terms of reducing the burden placed on them to remove content. 
From our interviews, it does not appear to be any clearer what is happening within the larger platforms and hosting providers, as seen in the discussion of the Statements of Reasons (SORs). One concern this creates is, if it is not clear whether reporting patterns are feeding back into algorithms, will it be clear how reporting is resulting in identifiable change on the platforms? Future work would consider how to bolster TF resolve, especially when considering that these organizations do not receive any additional funding for the potentially increased amount of reporting effort (and staffing) involved. Some participants spoke of taking on and training staff, but this is while there are also headlines of large platforms cutting their moderation teams; increases in reporting are outsourced but not resourced.

\paragraph{DSA resourcing and empowering user rights}
\label{sec:disc-empower}

Our findings suggest that Trusted Flaggers are positioned as a kind of `advanced' crowd-sourced reporter; they advance the causes they already supported and align therefore with their accreditation mandate (Art. 22 (2) DSA). The TF role is a mix of new capabilities to keep up with growing online content, and coordinated processes which -- when working well -- mean a fast-track reporting process with the platforms hosting content. 
These capabilities help TFs to `run to stand still', but with no additional resources. Referring to the DSA and the remit to recognize user rights to report illegal content, the TF apparatus does not enable any sort of `reporter mobility' where new causes emerge and are empowered by a TF role. Given the requirement for skills and time reported by our interviewees, some causes will gain TF representatives, and some will remain left behind as the upfront costs remain high and uncompensated. We expect the latter to include under-represented communities \cite{hasegawa2024weird} as before the existence of TFs, given that there is no additional resourcing.

\paragraph{The Trusted Flagger role and signaling legitimacy of user concerns}

Of great concern is whether under-represented causes will remain under-represented with the DSA, but now with a narrative that the DSA has empowered them (as if `job done' and move to the next concern, meaning users may be forgotten, as a potential unintended harm \cite{chua2019identifying} of the DSA). That TFs do not have additional resources makes a kind of sense within their remit and mandate, but requires clarity in communication as to what the DSA -- and specifically, the TF role -- is and is not for wider society. For instance, the DSA will seemingly not create any capacity to reach beyond a TF mandate to find marginalized user groups (i.e., those who experience ``pervasive negative treatment or exclusion'' \cite{matthews2025supporting}). The DSA will not suddenly, by its own existence, undo that `exclusion' in the specific sense of bringing onboard targeted reporting skills and a centralizing community hub such as a TF organization. That intent would still need to grow on its own, and we did not see that the TF role `freed up' energy, as registration and continued activity used up resources but in different ways, rather than making reporting more accessible. 

There is essentially a cap on TF capability, throttled by resources (of which there is nothing additional coming for TFs), so individual capability must grow somehow, or outreach capabilities under the DSA must grow. This relates to comments by P4 during their interview, that ``I know some organization even don't apply because they know that there won't be any financial compensation.''






\subsection{Recommendations}
\label{Recommendations}
Here we provide initial recommendations based on the outcomes of our interviews and workshop with Trusted Flaggers.\\

\noindent
\textbf{Standardized reporting processes.} Participants reported it being a drain to have to learn the ins and outs of sometimes very different reporting paths on different platforms. While this is not standardized, it also risks eating at TF resolve, if it continues to be unclear \textit{why} the reporting paths are different. This could turn out to be platforms also learning how best to implement DSA requirements, such as Art.~22(8) for guidelines and Art.~44(1)(a-c) for standards for reporting, design of submitting notices for users and TFs, where these guidelines and standards do not exist at the moment. However, it was not clear at this stage, whether platforms were indifferent to reporting being difficult for TFs to do. 

Such guidelines for example could also be provided by national regulators, as exemplified by Ireland's DSC \cite{coimisiun_na_mean_article_2024}.  
This need for standardization is also echoed in Art.~44(1)(c) DSA, providing the Commission with the regulatory power to set standards for ``electronic submission of notices by trusted flaggers under Article~22, including through application programming interfaces.'' Calls for standardized reporting and oversight channels align with proposals for ``legal compliance APIs" as a scalable enforcement interface \cite{goanta_case_2022}.\\

\noindent
\textbf{Meeting places for reporter organizations.} The workshop activity we conducted was explicitly requested by participants, and was well-attended. Trusted Flaggers would need such events to happen, if opportunities are needed to learn from each other and reduce up-skilling and reporting costs. 
Supporting TFs through national DSC-led events, cross-national through the Board (Art.~61), as well as better connection to the Commission, could show better coordination, sharing of best practices and more streamlined enforcement efforts for regulators on all levels. 
If there were no further effort by the wider enforcement apparatus, it may be that academic researchers can provide the meeting place (which would relate also to the need for objective measurement and audit, which is of academic interest too).\\

\noindent
\textbf{Do not lose users in keeping to mandates.} Trusted Flagger  participants spoke about their particular causes, and keeping to mandates defined around those causes. 
Further work should seek to understand the extent to which there are different kinds of TF, whether it be one more like a policing role keeping an eye on online content, or a support role to help users find support in a community of like-minded reporting entities.
In turn, future work will explore whether the mechanisms of the DSA make individual reporting actions any easier for the unfamiliar and unskilled user. 
Related to this is, if the mechanisms of the DSA support reduction of exclusion, but do not act to increase inclusion, then attention should be given to standardizing the reporting process; it may also be necessary to provide public guidance to users on what harmful content is, 
but also what is within their rights to do about it.\\

\noindent
\textbf{Act on the lack of knowledge about impacted users.} Recent research on at-risk user groups in privacy and security asks what we know about such groups, but also what we do not know \cite{wei2024sok}. If the TF role stays within a set mandate, then it is perhaps on researchers and general advocacy activity, to proactively understand which users or user groups we \textit{do not} know about, who want to participate in online discourse safely but do not (and do not know about reporting structures). 
This raises a question, as to whether every user should expect to be able to find a TF for their needs (be it specific or general), or may already get a signal of sorts from the ecosystem that their cause is not important, if no TF can be found which represents it (then already undermining their willingness to report, constituting a compound unintended harm \cite{chua2019identifying}). This phenomenon has been examined previously as the `likeness ideal', and related harms such as `exclusionary representativeness' \cite{chasalow2021representativeness}. Any `emerging' cause will need to develop TF capabilities quickly, and already know that they will be able to enact the role when registering with the DSC/Commission. A further question is whether 
an outreach and awareness capability is needed to reach under-represented users, for instance in a similar model to spyware clinics \cite{havron2019clinical} and privacy consultations for local communities \cite{rephrain2025}.

\section{Conclusion}
\label{conclusion}
In conclusion, our findings show that Trusted Flagger status increase the speed and likelihood of platform action, but the new role comes with the need for standardized processes, tooling, and costs, is constrained by fragmented reporting channels, and limited feedback from regulators and platforms. These operational frictions, coupled with insufficient activity capturing of TF's reporting in transparency artifacts (SOR) only insufficiently reflect lived practice, and create an accountability gap for evaluating TF work, platform responses, and its implications for user rights. Strengthening these TF mechanism therefore requires tailored standards for reporting illegal content, better platform cooperation, and a strategy that responds to the lack of additional resources to support the range of TF activities.


\bibliographystyle{ACM-Reference-Format}
\bibliography{refs}

\section{Generative AI Usage Statement}
In line with the ACM policy and FAccT guidelines, we used generative AI tools (including large language models) only for limited assistance with writing mechanics and formatting. Specifically, we used these tools to suggest grammar and fluency improvements to author-written text and to assist with LaTeX formatting tasks. We did not use generative AI to generate substantive content, arguments, results, analysis, or interpretations. All suggested edits were reviewed, selectively adopted, and verified by the authors, who remain fully responsible for the final manuscript.

\section{Ethical Considerations Statement}
\label{sec:ethics}
The study went through institutional Human Research Ethics Committee (HREC) review and received approval.
Consent for interviews and the workshop was obtained prior to the interviews.
Data was stored according to a data management plan (DMP) that was also reviewed and approved before HREC review. The DMP included several data protection measures, e.g., data minimization.

The interviews and online workshop were recorded. Participants were informed in advance that other participating organizations would be attending the online workshop. Multiple authors were present on the online workshop call, to take written notes in case any participants objected to recording, but no participant joining the online workshop objected.

Participants did not receive any financial compensation for taking part in the study, but instead received advance view of the analysis. For participants, one distinct benefit from the study was that the planned workshop met a request from several TFs, and it became a meeting point to share experiences. 
Most of the organizations involved in interviews also then accepted the conditions of participation and joined the Teams-based workshop call.

There is a risk that the participating organizations could be identified from the public list of registered Trusted Flaggers, given that -- at the time of writing -- it is a list of tens of organizations (rather than hundreds). The findings were written with this in mind, with consideration of how to convey meaningful detail while also being mindful of specific practices of any one organization. 

In preparation for the workshop, the summary of themes from the interview analysis (as in Appendix \ref{workshop_themes}) was reviewed by the authors several times, in an effort to limit any bias that may be introduced, so as not to sway the participants to make particular statements.

\appendix

\section{Interview Guideline}
\label{Interview_Guidelines}

\noindent
\textbf{Introduction}
\label{guidelines_intro}
\begin{enumerate}

    \item Can you explain your position and your work briefly?
    \item How does your work relate to users ['Recipients of the service' \emph{Art.~3(b) DSA}]? Which kinds of users or communities does your organization represent or support?
    \item Which of the work activities you have mentioned would you say fall within your Trusted Flagger role [\emph{Art.~22 DSA}]?
\end{enumerate}

\noindent
\textbf{Motivation / Process to Become a Trusted Flagger}
\label{guidelines_motivation}

\begin{enumerate}
    \item Have you been a Trusted Flagger before the DSA?
    \item Can you elaborate on the reasons to become a Trusted Flagger [\emph{Art.~22 DSA}]? (As opposed to conducting your activities without it; why did your organization decide to apply? [\emph{Rec.~62 DSA}])
    \item What specific qualifications did you provide in the process to become a Trusted Flagger [\emph{Art.~22(1) DSA}]?
\end{enumerate}

\noindent
\textbf{Function of Trusted Flaggers}
\label{function}

\begin{enumerate}
    \item What do you see as the main objectives for Trusted Flaggers [\emph{Art.~22(1)} in conjunction with \emph{Art.~16 DSA}]?
    \item How do organizations value or perceive the work of Trusted Flaggers?
\end{enumerate}

\noindent
\textbf{Reporting Process}
\label{Reporting_Process}

\begin{enumerate}
    \item What does your organization do as a Trusted Flagger [\emph{Art.~22(1)} and \emph{(3) DSA}]?
    \item What kind of content do you report? Is it largely within your area of expertise, or do you go beyond it [\emph{Art.~22(1), Art.~22(2)(a) DSA}]?
    \item What do you conceive as the role of average users in the Trusted Flagger system [\emph{Art.~3(b)} and \emph{Art.~22 DSA}]? How do users reach your organization, and what does ‘good’ user support look like from your perspective?
    \item Can you describe the tools you use to fulfill the requirements for Trusted Flaggers [\emph{Art.~22(3) DSA}]?
    \item Can you describe the reporting process for Trusted Flaggers in your experience [\emph{Art.~22(1), Art.~16 DSA}]?
    \item Who within your organization is responsible for carrying out Trusted Flagger tasks?
    \item What kinds of access controls or safeguards are in place for staff performing Trusted Flagger activities?
    \item Are you in contact with public authorities, and are you sharing your flagging activities with them [\emph{Art.~9,~10,~22 DSA}]?
\end{enumerate}

\noindent
\textbf{Interaction with Platforms}
\label{Interaction_Platforms}

\begin{enumerate}
    \item For which platforms do you conduct Trusted Flagger reporting tasks?
    \item In your opinion, how do platforms react to your flagging notifications [ \emph{Art.~22,~35(1)(g) DSA}] ?
    \item In your experience, how does prioritization play out in content moderation for your flags [\emph{Rec.~61}]?
    \item What kinds of interactions do you have with platforms, and how regular are they [\emph{Rec.~87}]?
    \item Are you reporting only for VLOPs and VLOSEs, or also for smaller online platforms? [\emph{Art.~16 and~22(1) DSA}]
\end{enumerate}

\noindent
\textbf{Legal vs Terms of Service}
\label{Legal_ToS}

\begin{enumerate}
    \item Do you only flag content under illegal content [\emph{Art.~3(h) DSA}] (legally relevant) reporting categories, or do you also flag [\emph{Art.~16 DSA}] under the Terms of Service of platforms [\emph{Art.~14 DSA}]?
\end{enumerate}

\noindent
\textbf{Statement of Reasons Data}
\label{SOR}

\begin{enumerate}
    \item Would you agree that the numbers indicated in the Statement of Reasons Database about Trusted Flagger notices accurately reflect your experience [\emph{Art.~17, Art.~22 DSA}]?
    \item (Optional) When looking at the Statement of Reasons data, how do you expect your organization’s activity—or your cause/community more broadly—to appear in those reports over time [\emph{Art.~17 DSA}]?
\end{enumerate}

\noindent
\textbf{Resources}
\label{Resources}

\begin{enumerate}
   \item How do you support the Trusted Flagger activity financially or in terms of resources [\emph{Art.~22 DSA}]?
    \item How many people are conducting Trusted Flagger tasks in your organization [\emph{Art. ~22 DSA}]?
    \item Can you describe the activities and amount of work that goes into completing a report [\emph{Art.~22(3) DSA}]?
    \item What happens to your records or stored data when a post you report is taken down by the platform [\emph{Art.~15(1)(b) DSA}]?
 \end{enumerate}

\noindent
\textbf{Website and Visibility}
\label{website}

\begin{enumerate}
    \item Could you describe how your website communicates your Trusted Flagger status and activities [\emph{Art.~22}]?
    \item Are there any specific features or elements on your website related to your Trusted Flagger work [\emph{Art.~22}]?
    \item Is there anything else you would like to mention or add that we haven’t covered?

\end{enumerate}

\newpage

\section{Screenshots of Trusted Flagger Activity Captured in the Statement of Reason Database}
\begin{figure}[ht]
    \centering
    \includegraphics[width=0.9\linewidth]{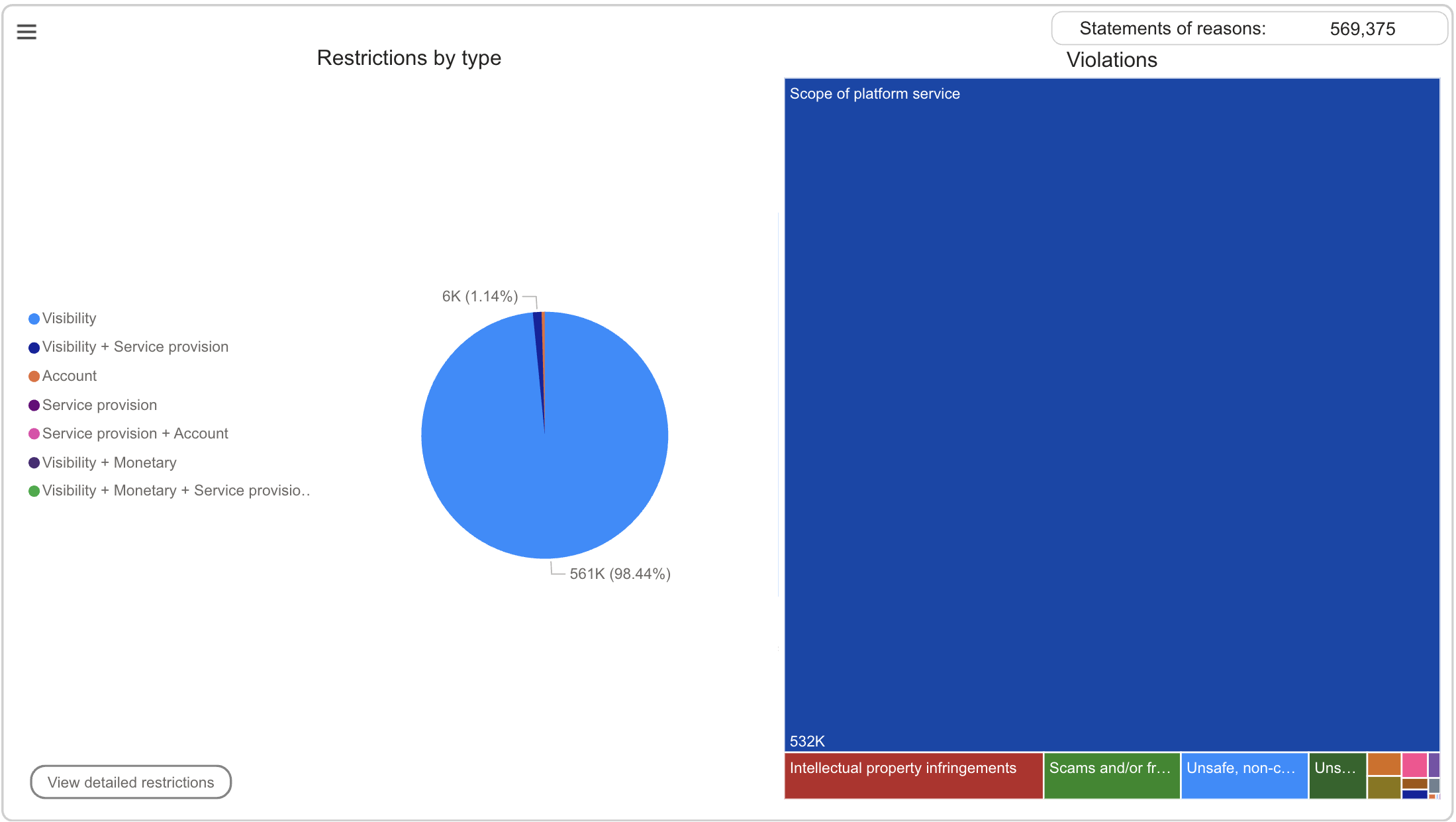}
    \caption{
    Statement of Reason Database Capturing Information About TF Notices 
    About Restriction Types and Violations by Statement of Reason Category Reason 
    taken on the 11/5/2025 \cite{european_commission_dashboard_2025, noauthor_api_nodate}.}
    \label{fig:SOR_1}
    \Description{Screenshot of an analytics dashboard. On the left, a pie chart titled “Restrictions by type” shows a large majority slice for “Visibility” (561K; 98.44\%) and small slivers for other restriction types. On the right, a tree-map titled “Violations” (569,375 Statements of Reasons) is dominated by a large rectangle labeled “Scope of platform service” (532K), with smaller rectangles for other violation categories.}
\end{figure}
\begin{figure}[ht]
    \centering
    \includegraphics[width=0.9\linewidth]{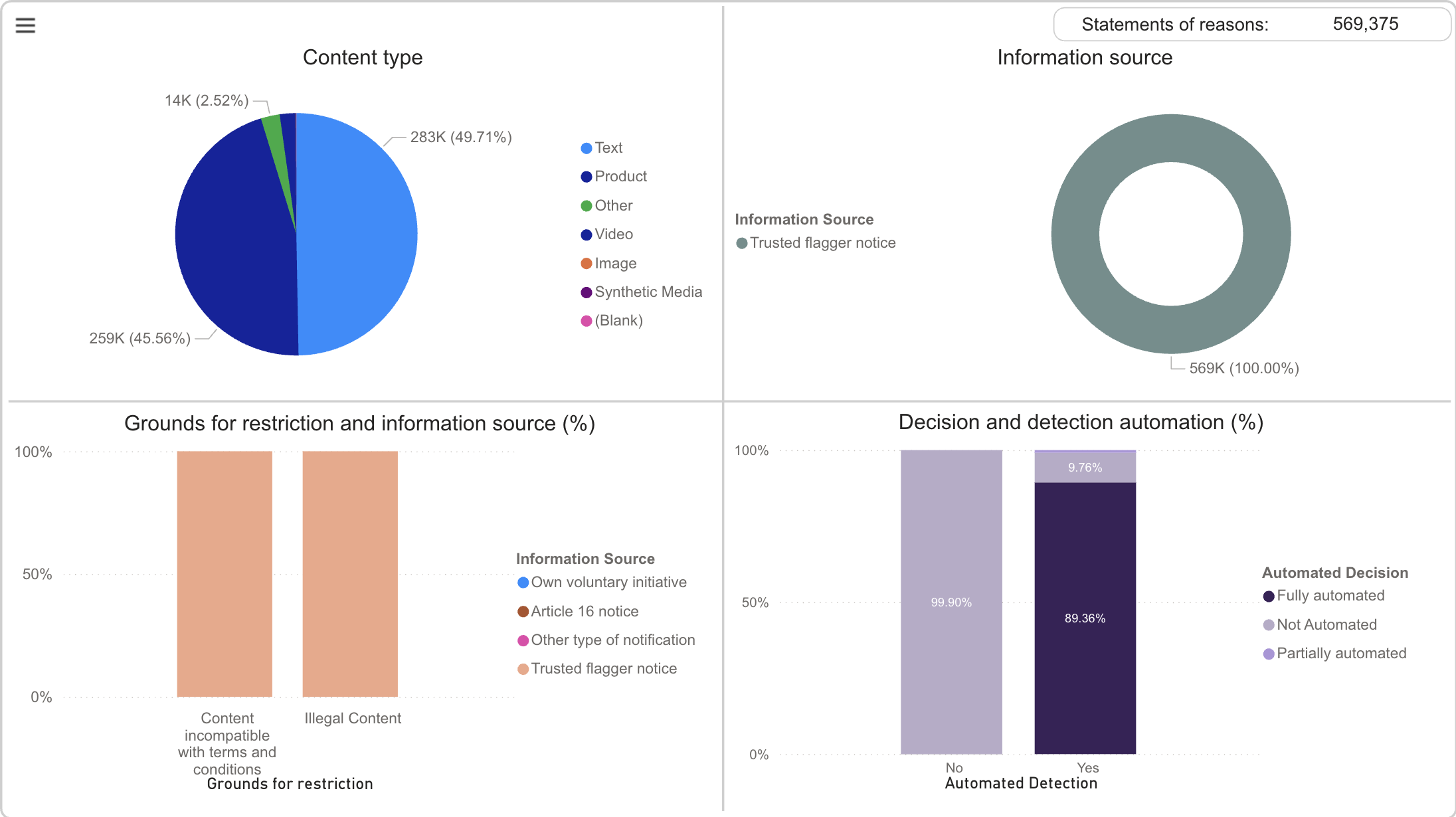}
    \caption{Statement of Reason Database Capturing Information About Content Type and Content Moderation Measures of TF Notices taken on the 11/5/2025  \cite{european_commission_dashboard_2025, noauthor_api_nodate}.}
     \label{fig:SOR_2}
    \Description{Screenshot of an analytics dashboard split into four panels. Top left: a pie chart titled ``Content type'' showing two large segments for Text (283K; 49.71\%) and Product (259K; 45.56\%), plus a small segment around 14K (2.52\%) and other minor categories listed in a legend. Top right: a doughnut chart titled ``Information source'' showing a single category, ``Trusted flagger notice'' (569K; 100.00\%). Bottom left: a 100\% stacked bar chart titled ``Grounds for restriction and information source (\%)'' with two bars (content incompatible with terms and conditions; illegal content), each entirely in the color corresponding to ``Trusted flagger notice.'' Bottom right: a chart titled ``Decision and detection automation (\%)'' comparing ``No automated detection'' (almost entirely ``Not automated'' at 99.90\%) versus ``Yes automated detection'' (mostly ``Fully automated'' at 89.36\% with smaller ``Not automated'' at 9.76\% and a thin ``Partially automated'' segment).}
\end{figure}
\newpage

\section{Main Themes Presented to Workshop participants}
\label{workshop_themes}

\begin{itemize}
    \item Platforms obstruct or fragment Trusted Flagger work
    \item Capacity, resourcing, and training constraints shape what TFs can do
    \item Regulatory coordination is uneven and (sometimes) difficult
    \item Platforms’ incentives and response patterns shape outcomes
    \item Illegality is operationally hard to pin down in practice
    \item Transparency and accountability artifacts can misrepresent TF labor and role
\end{itemize}

\end{document}